\documentclass[12pt]{iopart}

\usepackage{graphicx}
\begin{document}

\title[Coarse-graining network evolution dynamics.]{Coarse-graining
the dynamics of network evolution: the rise and fall of a networked society.}

\author{Andreas C. Tsoumanis$^{1}$,
Karthikeyan Rajendran$^{1}$,
Constantinos I. Siettos$^{2}$ and
Ioannis G. Kevrekidis$^{1,3}$
}

\address{$^{1}$ Department of Chemical and Biological Engineering, Princeton University, Princeton, NJ, USA\\
  $^{2}$ School of Applied Mathematics and Physical Sciences, NTUA, Athens, Greece\\
  $^{3}$ Program in Applied and Computational Mathematics, Princeton University, Princeton, NJ, USA}
\ead{atsouman@princeton.edu, krajendr@princeton.edu, ksiet@mail.ntua.gr and yannis@princeton.edu}

\begin{abstract}

We explore a systematic approach to studying the dynamics of evolving networks
at a coarse-grained, system level.
We emphasize the importance of finding good observables (network properties)
in terms of which coarse grained models can be developed.
We illustrate our approach through a particular social network model:
the ``rise and fall'' of a networked society \cite{Mars04rise}:
we implement our low-dimensional description computationally using
the {\em equation-free} approach and show how it can be used to (a) accelerate simulations
and (b) extract system-level stability/bifurcation information from the detailed dynamic model.
We discuss other system-level tasks that can be enabled through such
a computer-assisted coarse graining approach.

\end{abstract}

\maketitle

\section{Introduction}

The Erd\"{o}s-R\'{e}nyi random graph model \cite{Erd59random}, dating back to $1959$,
constitutes a landmark in the study of graphs.
There has been a renewed interest in networks (or graphs) in recent years
with more specific focus on complex emergent dynamics, spurred by discoveries
such as power-law degree distributions \cite{Falo99power-law,Yook02modeling},
small world behavior \cite{Watt98collective}, and more.
Dynamic network evolution models have been developed with an eye towards
constructing networks with specific structures/statistical properties;
an example is the preferential attachment mechanism \cite{Bara99emergence},
proposed as a source of scale-free structure in evolving networks.
Evolutionary network models are especially popular (and relevant) in the
social sciences \cite{Davi02emergence,Kump07emergence,Gros08robust,Zsch10homoclinic,Huep11adaptive-network}.
The focus in many of these models is on the correspondence between
local mechanisms (rules) of network formation and the resulting
large-scale ``system level" network structure and dynamics.
In this work, we explore a systematic approach to the computer-assisted study
of such models at the macroscopic, coarse-grained, {\em system} level
as opposed to the detailed, node-level, {\em microscopic} level.
The basic idea is to estimate the information necessary for coarse-grained computations
{\em on the fly}, using short bursts of appropriately designed detailed simulations.
Based on the selection of suitable coarse variables (observables), this
equation-free approach \cite{Kevr03equation-free,Kevr04equation-free:}
facilitates efficient computation at the coarse-grained level.
Accelerating coarse grained evolution computations, and/or enabling additional modeling
tasks such as coarse fixed point/bifurcation/stability computations
can significantly enhance our understanding of the system behavior at the macroscopic
level, even when an explicit coarse grained model is not available.
The approach may also help pinpoint collective network properties that are crucial
in the evolution process (and could thus be used as coarse variables).
The main aim of this paper is to demonstrate how our modeling approach to complex systems dynamics
has to be modified and extended in order to be usable in problems involving network evolution.
We will focus on a number of issues arising in this new context, like finding the
number of slow variables, constructing networks consistent with them, etc.

The paper is structured as follows:
In the next section we briefly describe an illustrative network evolution model
introduced by Marsili {\em et al} \cite{Mars04rise}.
We then discuss certain issues that arise in coarse variable selection,
and our chosen coarse-graining approach.
Results of the computational implementation of this approach are presented,
and we conclude
with a discussion of the scope and applicability of our approach,
as well as certain important open issues in the selection of coarse variables
for general network evolution problems.

\section{An illustrative case: the rise and fall of a networked society}

We revisit the model of the rise and fall of a networked society
\cite{Mars04rise} presented by Marsili {\em et al.} to illustrate our approach.
Their model exhibits complex emergent phenomena
arising from simple evolution rules; a brief
description of these rules follows:
Consider a (social) network consisting of $n$ agents (entities) involved in bilateral interactions.
The network is represented by $N = (V,E)$, where $V$ is the set of
vertices or nodes (agents) and $E$ is the set of edges or links, representing
interactions between agents.
In any time interval $[t,t+dt)$, the model evolves as follows:
\begin{enumerate}
\item Any existing edge $(i,j) \in E$ is removed with probability $\lambda dt$.
\item With independent probability $\eta dt$, every agent $i$ can form a link with (become a neighbor of) a randomly chosen
agent $j \neq i$.
Nothing occurs if agent $i$ is already a neighbor of (is connected with) agent $j$.
\item With independent probability $\xi dt$, every agent $i$ is allowed to ask a neighbor $j$ (randomly chosen)
to introduce him/her to a randomly chosen neighbor of $j$, say agent $k$.
If agent $i$ is not already connected
to agent $k$, a new edge is formed between $i$ and $k$.
Nothing occurs if agent $i$ is already a neighbor of (is connected with) agent $k$.
\end{enumerate}

\begin{figure}
\begin{center}
\resizebox{0.54\columnwidth}{!}{%
  \includegraphics{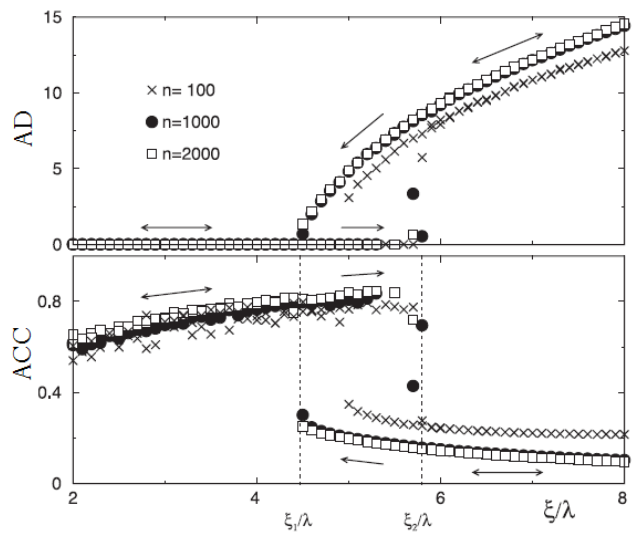}
}
%\vspace{5cm}
\caption{
Bifurcation diagram obtained using direct temporal simulations, reproduced
from \cite{Mars04rise} with permission (copyright $2004$ National Academy of Sciences, U.S.A).
Steady state values of average degree, average clustering
coefficient and giant component size are plotted against the ratio
of parameters, $\xi/\lambda$.
$\eta$ and $\lambda$ values were fixed at $0.001$ and $0.1$ respectively.
Notice the robustness of the results to the network size.
}
\label{fig:BD}
\end{center}
\end{figure}

A detailed description of the motivation behind the model and the richness of the resulting dynamics is given
in \cite{Mars04rise}.
A numerical bifurcation diagram is also obtained there through
extensive direct temporal simulations of the model in different
parameter regimes.
This bifurcation diagram is reported again in Figure \ref{fig:BD},
where the average degree\footnote{The degree of a node in a network is the number of edges connected to the node.}
(AD) and the average clustering coefficient\footnote{Clustering coefficient of a node in a network is the ratio of
the number of triangles associated with the node to the maximum possible number of triangles
that could be associated with that node given its degree.}
(ACC) are plotted against a bifurcation parameter (the ratio $\xi/\lambda$).
The parameters $\xi$ and $\eta$ represent the rates at which the
agents form new connections through their existing friends and
through random sampling of the network, respectively; $\lambda$ is
the rate at which the connections between agents disappear.
To produce this bifurcation plot, $\eta$ and $\lambda$ were fixed at
$0.001$ and $0.1$, respectively.
The figure suggests generic hysteresis behavior as the parameter
$\xi$ is varied.
When $\xi \in [\xi_1,\xi_2]$ ($\xi_1$ and $\xi_2$ are marked in Figure \ref{fig:BD}), the system can
reach two distinct stable steady states depending on the initial condition: (a) a highly connected state with low
clustering and (b) a sparsely connected state with high levels of clustering (localization).
From dynamical systems theory, one anticipates a branch of unstable
steady states (not obtainable by direct simulations) ``connecting''
(in a coarse-grained bifurcation diagram) these two stable branches.
In fact, a mean field approximation which qualitatively reproduces
these characteristics of the full model behavior (including the
unstable branch) was also reported in \cite{Mars04rise}.

\section{Selecting an appropriate low-dimensional description}

One of the crucial steps in developing a reduced, coarse-grained description of any
complex system is the selection of
suitable observables (coarse variables).
We borrow simple principles from dynamical systems to help us in this pursuit and to provide
the rationale corroborating our coarse variable choices.
In order for a system to exhibit low-dimensional behavior, one expects
a separation of  time-scales to prevail in the evolution of different
variables in the system phase space.
The basic picture is that
low-dimensional subspaces
({\em slow manifolds}, parameterized by ``slow'' variables)
contain the long-term dynamics,
while fast evolution in the transverse directions (in the ``fast'' variables) quickly brings the
system trajectories close to the (attracting) slow manifolds.
%
%The long-term dynamics of the system are described solely by the state %of the system in the
%lower-dimensional subspace called the {\em slow manifold}, which is %spanned by
%these slow directions.
%
Thus, the problem of selecting good coarse variables is translated to the problem of finding a set of
variables that successfully parameterize the slow manifold(s), {\em when such a manifold exists}.

\begin{figure}
\begin{center}
\resizebox{0.72\columnwidth}{!}{%
  \includegraphics{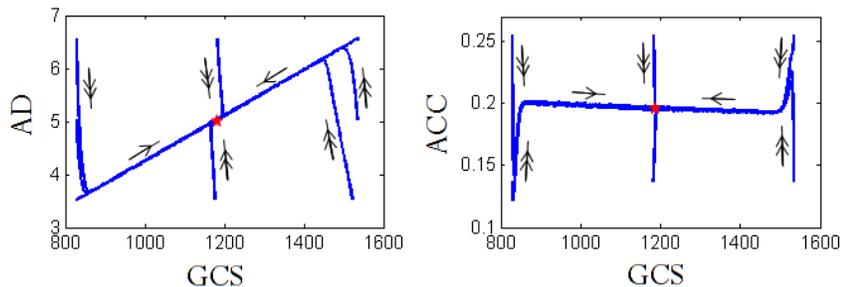}
}
%\vspace{5cm}
\caption{
Phase portrait in terms of giant component size (GCS) and average degree (AD) (left) or
average clustering coefficient (ACC) (right) showing transients from a variety of initial conditions.
Fast (resp. slow) temporal evolution is indicated by a double (resp. a single) arrow.
}
\label{fig:PP}
\end{center}
\end{figure}

In order to search for such a slow manifold in the context of the rise and fall model,
we explore the relevant state space through direct simulations as follows.
We first compute the steady (stationary) state of the model (for a fixed set of parameter values)
by direct simulations, and evaluate a few network properties at this stationary state.
We focus on three typically studied network properties: the giant component
size\footnote{The giant component size of a network is defined as the number of nodes in the largest
connected component of the network.} (GCS), the average degree (AD) and the average clustering coefficient (ACC).

\begin{figure}
\begin{center}
\resizebox{0.9\columnwidth}{!}{%
  \includegraphics{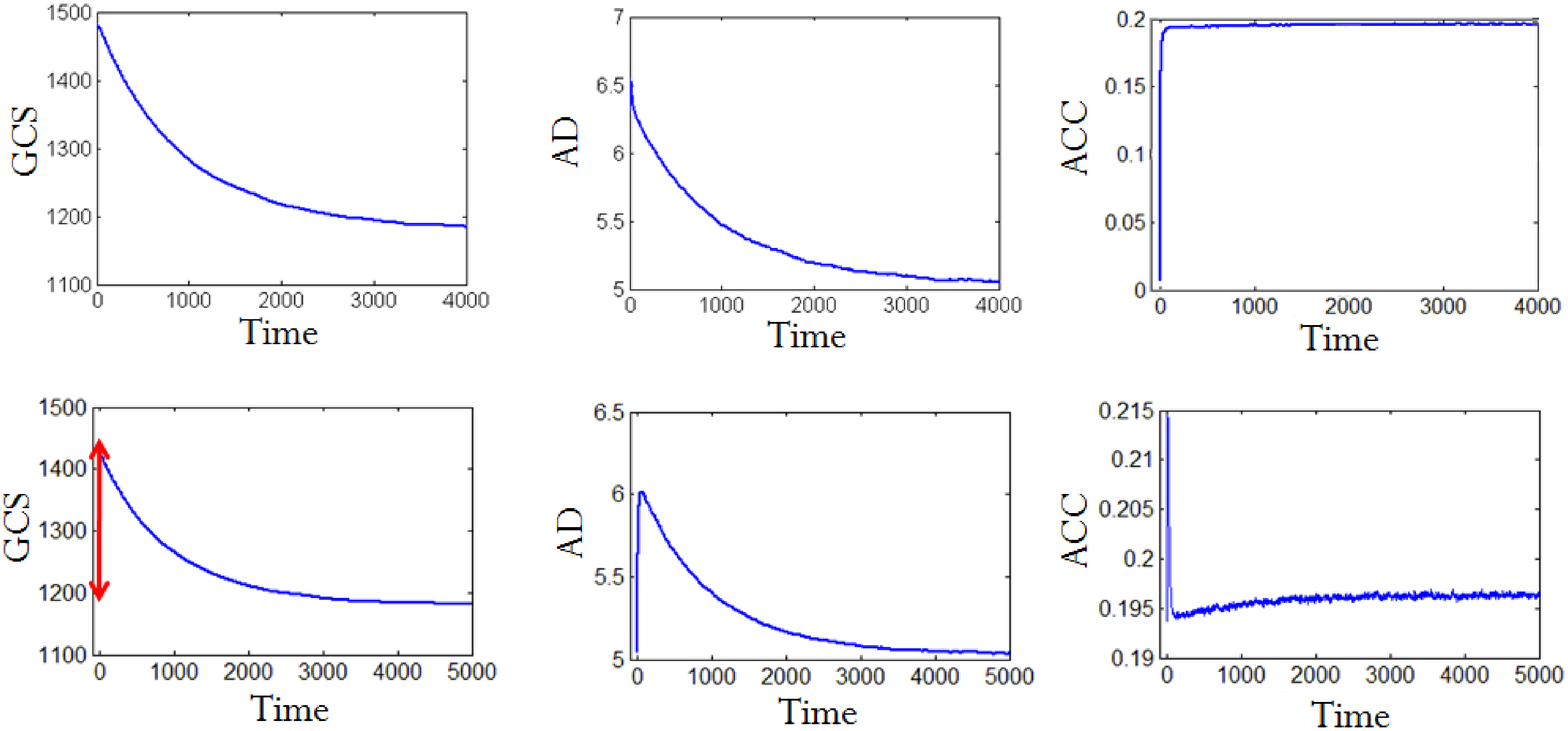}
}
%\vspace{5cm}
\caption{ Evolution of certain (expected) graph properties: average
degree, average clustering coefficient and giant component size.
The top panels correspond to evolution from random initial networks, while the initial graphs
for the bottom panels were obtained by creating networks exhibiting the stationary
degree distribution and the stationary value of ACC.
The red arrow indicates the magnitude of our perturbation away from the stationary
GCS value.
}
\label{fig:Per1}
\end{center}
\end{figure}

For our simulations, the parameters of the model are taken to be $n=2000$, $\lambda = 0.1$, $\eta = 0.001$ and $\xi=0.5$.
Fig.~\ref{fig:PP} depicts the phase portrait of AD and ACC versus
GCS starting from several different initial conditions.
A one-dimensional slow manifold parameterized by the giant
component size is clearly suggested by the simulations.
The time evolution of these network properties from a random initial condition is shown in the top
panels of Fig.~\ref{fig:Per1}.
It should be noted that, in order to observe smooth trajectories representative of the {\em expected}
evolution, ensembles of several realizations --typically $100$-- are averaged to create these plots.
The trajectories suggest that the giant component size and the
average degree appear to vary slowly over time, while the average
clustering coefficient has a fast initial transient followed by a
slower evolution to the steady state.

In order to numerically determine the directions in state space along which the system evolves slowly,
we then perturb the system away from the stationary state itself, by varying different
sets of network properties, and observe the evolution of the properties of the resulting perturbed networks.
In one such perturbation experiment, we construct networks {\em with the stationary
degree distribution and average clustering coefficient} and
perturb the system away from the stationary value of giant component size.
We then let the system evolve naturally, and observe, in Fig.~\ref{fig:Per1}(bottom), the
evolution of the three above-mentioned network properties.
The perturbed giant component size varies slowly from its initial value
and eventually approaches steady state.
The evolution of the average degree and the average clustering coefficient
show fast initial transients followed by a slow approach to steady state.
These observations suggest that the giant component size may be a ``good" slowly evolving variable,
to which the average degree and the average clustering coefficient
(degree and triangle information) become quickly slaved
(and remain slaved during long-time evolution).

\begin{figure}
\begin{center}
\resizebox{0.9\columnwidth}{!}{%
  \includegraphics{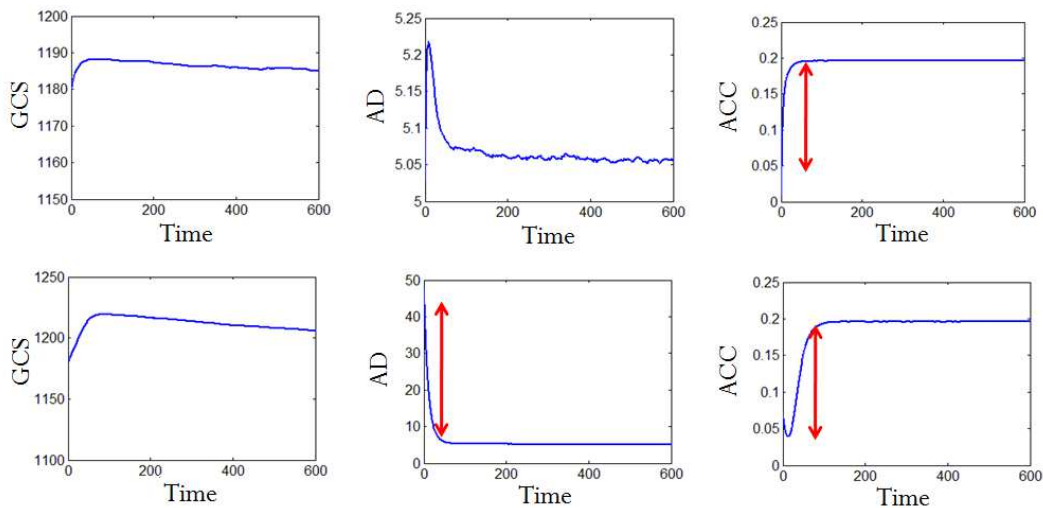}
}
%\vspace{5cm}
\caption{ Evolution of certain (explected) graph properties: average
degree, average clustering coefficient and giant component size.
The top panels correspond to evolution from initial graphs exhibiting both the
stationary degree distribution and the stationary giant component size.
The initial graphs for the evolution shown in the bottom panels were constructed to exhibit
{\em only} the stationary giant component size value.
The (red) arrow(s) in each row indicate the magnitude of our
perturbation away from the stationary property value(s). }
\label{fig:Per2}
\end{center}
\end{figure}

We perform a number of similar numerical perturbation computations that, taken together, reinforce
the view that the size of the giant component (a single scalar variable) may be
a suitable candidate coarse variable.
%to describe (parameterize) the slow manifold observed in
%the long-term dynamics of the problem.
%
A couple of such computations supporting this idea are shown in Fig.~\ref{fig:Per2}.
The top panels show the temporal evolution of (the expected value of) a few network properties
from initial networks constructed to exhibit {\em both the
stationary giant component size and the stationary degree distribution}.
We repeat these computational experiments initializing the evolution from networks exhibiting
{\em the stationary value of the giant component size only}\footnote{A network with a given value of giant component size
was constructed by creating an Erd\"{o}s-R\'{e}nyi random graph whose number of nodes equal the size of
the giant component. The Erd\"{o}s-R\'{e}nyi graphs that we create have an average degree
around $8$; if the desired giant component size is less than $9$, then we do create a clique.
We then add nodes of zero degree
until the network has the prescribed total number of nodes.}
(and not exhibiting the stationary degree distribution).
The corresponding evolution results are shown in the bottom panels of Fig.~\ref{fig:Per2}.
We find that, if we initialize with networks that exhibit the
``correct" stationary value of the giant component size, the average
degree will quickly evolve to the neighborhood of its stationary
value -``come down" on the slow manifold- and will then slowly
approach it on the same time scale as the giant component size does.
This would suggest that the average degree is a ``fast variable".
%that takes
%it close to the steady state value (from where it has a slow evolution)
%whether the initial network exhibited the stationary value of the
%average degree or not.
%
By the same type of argument, Figs.~\ref{fig:Per1} and \ref{fig:Per2} suggest that the
average clustering coefficient, also, is a fast variable and hence does not need to be
explicitly included in a model for the long-term evolution.

In both cases shown in Fig.~\ref{fig:Per2}, the giant component size itself
has a fast initial evolution window that
takes it momentarily away from the steady state value,
and then it slowly evolves back towards it ({\em on} the slow manifold).
While the giant component size is a good candidate coarse variable
to parameterize the slow manifold (and the long-term dynamics),
these transients indicate that it is not a {\em pure} slow variable
-- its evolution appears to have initial fast as well as slow
components.
A clear illustration (and explanation) of this behavior,
and of the concept of a ``pure" slow variable in the context of a simple singularly perturbed problem
can be found in the Appendix.
These dynamics imply that initializing with a desired value
of the giant component size is not sufficient; additional
care must be taken to find a network that exhibits
this value {\em but also lies close to the slow manifold}
(explained later below).
If the variable was a {\em pure} slow one, then just
a few simulation steps would guarantee the latter property (lying close to
the slow manifold, see
again the simple illustration in the Appendix).

\section{A coarse-grained model}

Once we choose suitable coarse variables, computations
at the coarse-grained level are carried out using the {\em equation-free} framework
\cite{Kevr03equation-free,Kevr04equation-free:}.
This framework is used to accelerate simulation and also enable the performance
of a number of additional tasks (such as fixed point
computation and stability analysis) at the coarse-grained level.
In this approach, computations involving the coarse variables are performed with the help of suitably defined operators which translate between coarse and fine descriptions.
Short bursts of fine scale simulation followed by
observation and post-processing of the results at
the coarse scale enable on-demand estimation of numerical
quantities (residuals, actions of Jacobians, time-derivatives) required for coarse numerics.
The operator that transforms coarse variables to the detailed, fine variables is called the {\em lifting operator}, $L$,
while the operator that transforms fine variables back to coarse variables is called the {\em restriction operator}, $R$.
One can thus evolve the coarse variables of a system forward in time for a given number of steps $t$ by performing
the following operations:
\begin{enumerate}
\item \textbf{Lifting ($L$)}: Find a set of detailed variables (networks) consistent with the initial value(s) of the coarse variable(s) (here, the giant component size).
\item \textbf{Microscopic evolution}: Evolve the fine variables (the nodes and edges of the networks) for a specified time $t$ using the detailed, microscopic
evolution rules of the system.
\item \textbf{Restriction ($R$)}: Observe the fine variables, from the final stage(s) of the previous step, at the coarse scale.
\end{enumerate}
These steps constitute what is known as the {\em coarse time-stepper}, which acts as a substitute code for the
unavailable macroscopic evolution equations of the system.
In terms of the lifting and restriction operators, the {\em coarse time-stepper} $\Phi_t$ can be written in terms of the {\em fine scale} evolution operator ($\phi_t$) as:
\begin{equation}
\Phi_t(\cdot) = R \circ \phi_t \circ L(\cdot).
\label{eq:cts}
\end{equation}

Using this coarse time-stepper in the form of a black-box subroutine, we can ``wrap'' around it
a number of different algorithms (initial value solvers, fixed point solvers, eigensolvers) that perform system-level tasks.
In our present example, the microscopic description of system evolution is the model itself, defined in
terms of the network structure (fine variables, information about the edges between nodes), while
the giant component size is the single coarse variable.
Hence, the {\em restriction} operation consists of simply evaluating the giant component size of a given network.
The {\em lifting} operation, however, consists of constructing a network {\em with a specified value of giant component size};
this by itself, however, is not sufficient for EF computations, as we need a network that has the specified GCS
{\em and} lies on or close to the slow manifold.

We show a schematic representing the state space of our dynamical system in Fig.~\ref{fig:LiftS}.
The x-axis represents our coarse variable, the giant component size (GCS),
while the y-axis represents the directions corresponding to
all other possible variations in the network that do not alter the GCS.
The solid (blue) line denotes the slow manifold, which is drawn in a manner
suggesting a good one-to-one correspondence with our chosen coarse variable (GCS).
The vertical, dashed (black) line represents a family of graphs
having the same (prescribed) giant component size.
Let's pick one graph in this family denoted by the (green) point number $1$
and use it as an initial condition in the model we study.
If the giant component size were truly a slow variable, the system
would ``quickly'' evolve to a point on the manifold {\em with a giant component size very similar to the
starting value}.
However, as we observed before, the giant component size is {\em not a pure slow variable}:
as the system dynamics quickly evolves from point $1$ to
point $2$, we do approach the slow manifold, yet our
giant component size has significantly changed.
It is therefore important that our lifting operation
constructs networks that not only conform to the
prescribed giant component size but
also lie {\em close to the slow manifold}.
In terms of our schematic caricature of Fig.~\ref{fig:LiftS},
if the vertical line corresponds to the required value of GCS and
the lifting operation is required to produce a graph close to point $D$.
The issue of initializing on the slow manifold is a crucial one in many scientific
contexts that involve model reduction (e.g. in meteorology, \cite{Lore86on}) and algorithms
for accomplishing it in a dynamical systems context are the subject of current research
(see e.g. \cite{Gear05constraint,Zaga12stability}).

\subsection{Our Lifting Operator}

\begin{figure}
\begin{center}
\resizebox{0.54\columnwidth}{!}{%
  \includegraphics{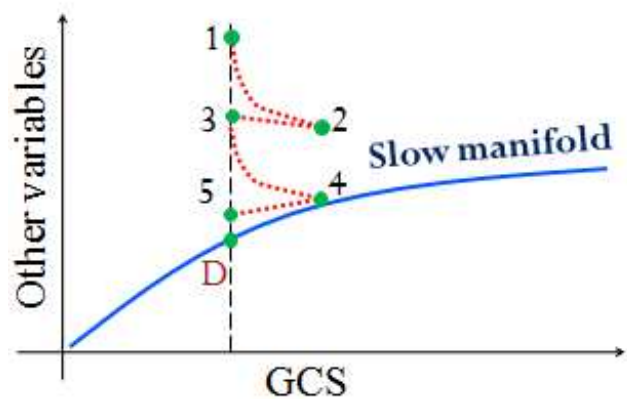}
}
%\vspace{5cm}
\caption{
Schematic of our lifting procedure to create graphs with a specified
giant component size {\em which also lie close to the slow manifold}.
The process begins with the network labeled $1$.
}
\label{fig:LiftS}
\end{center}
\end{figure}

\begin{figure}
\begin{center}
\resizebox{0.9\columnwidth}{!}{%
  \includegraphics{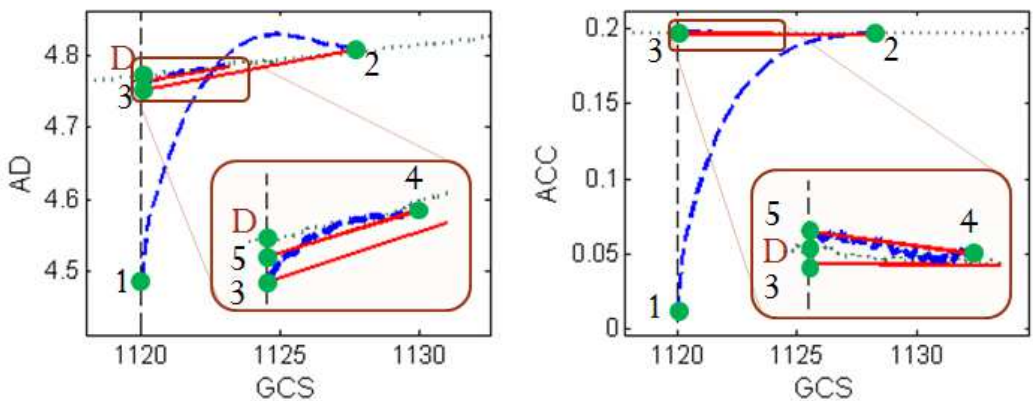}
}
%\vspace{5cm}
\caption{
Graph evolution during our lifting operation is tracked through plots of average degree (AD)
and average clustering coefficient (ACS) versus giant component size (GCS).
Here, the required value of giant component size is $1120$.
The graph with this desired value of giant component size \emph{and} on the
slow manifold (plotted as thin, dotted green line) is labeled as
`D'.
The lifting operator arrives at it through the intermediate graphs $1$ to $5$, in that order.
The dashed (blue) lines denote evolution of the dynamical model:
this is the ``process step".
The solid (red) lines denote corrections for the change in giant
component size: this is the ``adjustment step".
The combination of two such steps constitutes a single full iteration.
Two such full iterations are shown in this illustration.
}
\label{fig:Lifting}
\end{center}
\end{figure}

We implement such a lifting operation beginning with a network that possesses the required value of giant component size, by creating an
Erd\"{o}s-R\'{e}nyi random graph\footnote{Note that these initial graphs are essentially initial
guesses for our lifting operator, which creates graphs having the specified value of
giant component size \emph{and} lying close to the slow manifold.
A good initial guess reduces the computational effort involved in finding such a graph.
Thus, although we can successfully lift from a variety of reasonable different
initial conditions, the initial condition does matter for computational efficiency.}
whose number of nodes is equal to the size of the giant component;
we then add nodes of zero degree until the network has the correct number of total nodes.
Let us denote this result by point $1$ in Fig.~\ref{fig:LiftS}.
We then run the model for a few (here, typically $70$) time steps and obtain network $2$,
which lies close to the slow manifold, but has a different value of GCS than required.
We now appropriately add/remove enough nodes from this network $2$ until the resulting
network $3$ has the required number of nodes in its giant component.
This adjustment step is illustrated as a straight line segment in
the schematic, for convenience (we do not control the other 
variables during the adjustment step).
If we add nodes to the giant component, we assign them a degree sampled from the current giant component
degree distribution.
This can be done by randomly selecting a node in the current giant component and using its degree as the degree of the added node.
This node is then connected to as many nodes of the giant component as its degree.
If we remove nodes from the giant component, they become isolated nodes of degree $0$.
When removing nodes, one must of course be careful not to break up the giant component.
This is taken care of by first constructing a spanning tree of the giant component, and only removing the fringe nodes of degree $1$ {\em in this tree}.
All the edges connected to this node in the original graph are then finally removed.
In this paper these lifting steps are repeated two or three times as necessary ($2 \rightarrow 3, 4 \rightarrow 5,\ldots$)
so that we obtain a network close to the desired graph $D$ in the schematic.
We also show an actual sample graph evolution during our lifting operation in Fig.~\ref{fig:Lifting}.
In this case, the required giant component size is $1120$.
We start from an Erd\"{o}s-R\'{e}nyi random graph (average degree around $4.5$) and
perform two iterations of the lifting operation.
As explained earlier, each iteration consists of a process step (dashed, blue line)
and an adjustment step (solid, red line).
The desired, ``lifted" graph is labeled `D'; the lifting operator
arrives at it through intermediate graphs in the sequence from $1$
to $5$.

\begin{figure}
\begin{center}
\resizebox{0.45\columnwidth}{!}{%
  \includegraphics{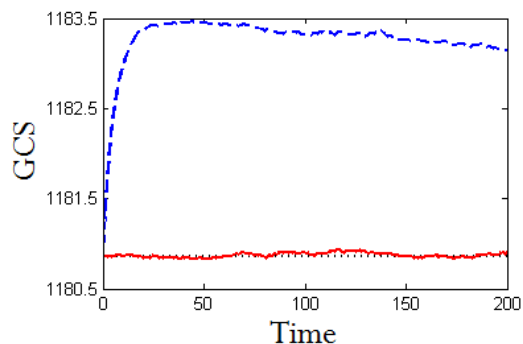}
}
%\vspace{5cm}
\caption{
Trajectories of giant component size observed by simulations from
two different classes of initial conditions.
The dashed (blue) curve corresponds to the case where the initial condition
is a random network created with the stationary values of the giant component size and
of the average clustering coefficient as well as the stationary degree distribution.
The solid (red) curve corresponds to initial networks created by our lifting procedure,
exhibiting the stationary GCS {\em and} lying on/close to the slow manifold.
}
\label{fig:Lift}
\end{center}
\end{figure}

We illustrate the effect of this lifting procedure by evolving the model from
two different initial conditions.
In the first case, we use a random network created to
exhibit the stationary values of the
giant component size, the degree distribution and the average clustering coefficient.
The evolution of GCS in this case, shown in Fig.~\ref{fig:Lift} as a
dashed line (blue), is reminiscent of the left panels of
Fig.~\ref{fig:Per2}.
Thus, it is clear that initializing with the stationary values of GCS, ACC and degree distribution
{\em is not sufficient} to keep the system close to the slow dynamics; fast transients that take these variables
away from their stationary values ensue.

In the second case, we use the lifting procedure just described above to create networks
with the stationary GCS value {\em that also are close to the slow manifold}.
Fig.~\ref{fig:Lift} shows the evolution of the GCS when we run the model starting from these
lifted networks as a solid (red) line.
Now the giant component size remains close to its stationary value,
as desired.
This suggests that our lifting procedure is successful.

%
%For different choices of coarse variables, the lifting
%step could imply
%creating networks with specified values of
%different properties (see, e.g. \cite%{Goun11generation}).
%

\section{Computational results and discussion}

\begin{figure}
\begin{center}
\resizebox{0.7\columnwidth}{!}{%
  \includegraphics{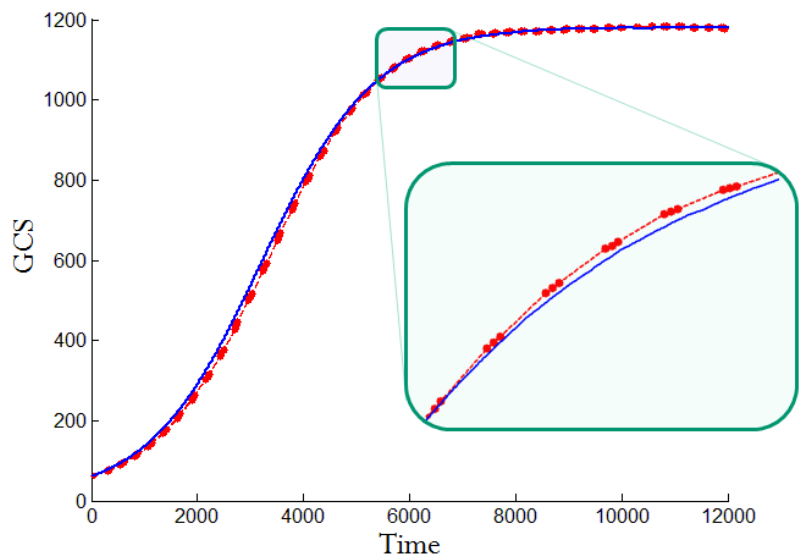}
}
%\vspace{5cm}
\caption{
Coarse projective integration of the network evolution, shown by the dotted (red) line, is performed
by running direct simulations for $60$ time steps and then projecting forward by $210$
time steps, lifting, and then repeating the procedure.
The circle markers correspond to portions of time when the system is actually evolved using fine scale direct
simulations.
{\em Full} direct simulations are also shown as a solid (blue) line for comparison.
}
\label{fig:Res}
\end{center}
\end{figure}

We first validate our coarse-grained modeling by illustrating
{\em coarse projective integration} of the (expected,
averaged over many realizations) system dynamics; we then present coarse bifurcation computations.
Let $g$ denote the coarse variable, the giant component size.
We start with an ensemble of $2000$ realizations of networks (each with $n=2000$ nodes) exhibiting a specified initial
value of this giant component size $g_0$.
$g$ is evolved using the coarse time-stepper (Eq.~\ref{eq:cts}) for a few (here, 60) time steps.
The giant component size is observed from simulations and its time derivative is {\em estimated}
from the last few time steps of the simulation.
This information is used to ``project'' the coarse variable
forward in time through a specified (``long'') time horizon.
At the projected time we {\em lift} - that is, we construct networks that exhibit the {\em predicted} value of
giant component size {\em that also lie on the slow manifold} as described in the previous section.
We used simulation and projection time horizons of $60$ time steps
and $210$ time steps, respectively.
The time-derivatives required for projection were estimated by using the observed values of giant component size at
times $50$ and $60$ of the current simulation step.
We show a trajectory computed through coarse projective integration in the left plot
of Figure \ref{fig:Res} as a dotted (red) line.
The data points corresponding to the simulation times are marked as filled (red) circles.
The trajectory from full direct simulation is plotted as a solid (blue) line for comparison.
It is clear that the two evolutions are visually comparable; yet the projective computation only necessitated simulations for $\frac{60}{270}=22.2\%$, roughly one-fifth of the time.
These savings came because of the {\em slow} evolution of the (expected value of the)
coarse variable, as constrasted to the fast, stochastic (binary - no smooth time derivatives!)
evolution of the individual realization fine scale variables.
The actual computational savings should also take into account the cost of repeated lifting
operations (here, the actual composite computational savings were at $35\%$).

\begin{figure}
\begin{center}
\resizebox{0.6\columnwidth}{!}{%
  \includegraphics{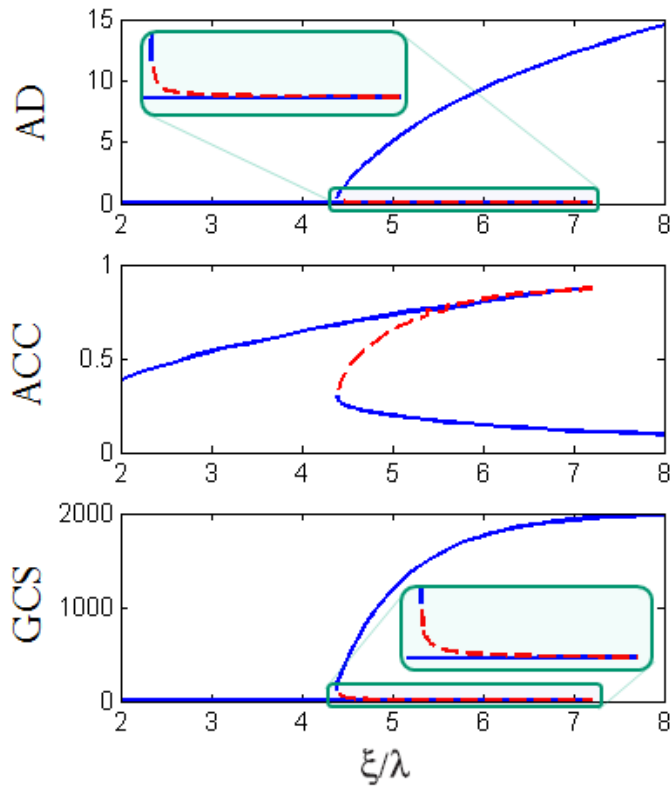}
}
%\vspace{5cm}
\caption{
Bifurcation diagrams obtained by coarse fixed point computations.
Steady state values of average degree, average clustering
coefficient and giant component size are plotted against the ratio
of parameters, $\xi/\lambda$.
The stable branches are plotted as solid (blue) lines, while the unstable branches are
plotted as dotted (red) lines.
$\eta$ and $\lambda$ values were fixed at $0.001$ and $0.1$ respectively.
}
\label{fig:ResBD}
\end{center}
\end{figure}

Next, we find the coarse steady state of the process time-stepper by solving the
following equation (as opposed to finding it by simply evolving the model in time):

\begin{equation}
F(g) = g - \Phi_{60}(g).
\label{eq:sse}
\end{equation}

\noindent
This procedure has the advantage of being able to find stable {\em as well
as unstable} steady states; the latter cannot be found using simple direct simulation.
The coarse steady solution, $g_s$, to Eq.~\ref{eq:sse} can be found via a Newton-Raphson procedure.
The Jacobian (derivative of $F(g)$ with respect to $g$) required to perform each Newton-Raphson iteration is
estimated numerically by evaluating F at neighboring values of $g$.
Since this problem involves a single coarse variable, the linearization consists of a single element,
that is easy to estimate through numerical derivatives.
For problems with large numbers of coarse variables estimating
the many components of the linearizations becomes cumbersome,
and methods of matrix-free iterative linear algebra (like GMRES
as part of a Newton-Krylov GMRES, \cite{Kell95iterative})
become the required tools.

We compute the coarse steady states over a range of parameter values, corresponding to the
bifurcation diagram shown in Figure \ref{fig:BD}.
We also keep track of complete network structures corresponding to
the coarse steady states on the different solution branches, and
store information about steady state values of average degree and
average clustering coefficient also.
Views of the bifurcation diagram thus produced are shown in the right panels of Figure \ref{fig:Res}.
The bifurcation results qualitatively (and visually quantitatively) agree
with those obtained in \cite{Mars04rise} as reproduced in
Figure \ref{fig:BD}; unstable branches of the bifurcation diagram
have now been recovered, and the instabilities (the boundaries of
hysteresis) are confirmed (as expected) to be coarse saddle-node bifurcations.

\begin{figure}[t]
\begin{center}
\resizebox{0.45\columnwidth}{!}{%
  \includegraphics{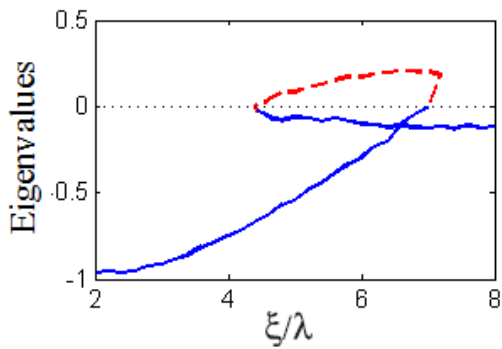}
}
%\vspace{5cm}
\caption{
Eigenvalues of the linearization of the coarse time-stepper with respect to the
coarse variable (the giant component size).
The portion of the curve above the x-axis corresponds to the unstable middle branch
(red, dotted lines) in the bifurcation diagrams of Fig.~\ref{fig:Res}.
The stable branches are plotted as solid (blue) lines.
}
\label{fig:ev}
\end{center}
\end{figure}

In addition to the steady states, the eigenvalues of the linearization of the coarse time-stepper
with respect to the coarse variable (the giant component size) were also estimated at the located steady states.
These eigenvalues, which give quantitative information about the stability of the different
bifurcation branches, are plotted in Fig.~\ref{fig:ev}.
The portion of the curve above the x-axis corresponds to the middle branch
in the bifurcation diagrams of Fig.~\ref{fig:Res}.
Thus, the upper and lower branches in the bifurcation are found out to be stable
branches, while the middle branch is unstable as anticipated.
Increased noise in the simulations close to the right turning point renders
the diagram there slightly imperfect.
Once again, the problem has a single-component linearization, which upon convergence
of the Newton iteration also constitutes an estimate of the relevant eigenvalue.
More generally, matrix-free Krylov-based Arnoldi methods must be used for
the estimation of the leading coarse eigenvalues
(see e.g. \cite{Arpa99lapack} and \cite{Siet03enabling}).

\section{Conclusions}

In this work we have presented a computer-assisted framework for the systematic coarse-graining of the dynamics
of evolutionary network problems.
For illustration purposes, we used a simple model (introduced by Marsili {\em et al.} in
\cite{Mars04rise}), which nevertheless results in the emergence of complex dynamics at the
coarse-grained level, including multiplicity of coarse-grained stationary states.
The analysis is relatively simple, yet it demonstrates the algorithmic tasks that one can attempt using
the proposed approach.
Coarse projective integration, that is, acceleration of the detailed evolutionary
network model dynamics, as well as coarse fixed point and coarse stability computations
can now be performed using the detailed evolutionary
network model as a black box simulator.
This is accomplished by circumventing the need for obtaining {\em explicit}, closed-form macroscopic analytical approximations,
effectively providing computer-assisted ``on-demand'' model reduction.
The key assumption here is that evolutionary equations at the macroscopic level can,  {\it in principle} be obtained,
but the required closures are not available
due to the complexity of the underlying fine scale dynamics.

The illustrative model also serves to highlight some important issues that arise
in this new context of adapting our coarse-graining framework to problems involving
complex network dynamics.
The selection of good coarse variables is always an important and often a challenging step.
For this illustrative model, we established through numerical experimentation that the giant
component size is a good coarse variable to capture the slow dynamics,
as other network statistics have predominantly fast dynamics and get quickly slaved to the slow dynamics.
We also created a procedure for initializing networks (our "lifting") consistent with slow dynamics.
By computationally implementing our model reduction, we recovered the entire bifurcation diagram
with this one coarse variable, which suggests that this "effective one-dimensionality" holds over
the entire parameter range we have studied.
We should add that, within the equation-free framework,
there is a number of additional tasks that can in principle be performed,
such as the implementation of algorithms that converge on, and continue,
higher codimension bifurcation loci, or also algorithms that design
stabilizing controllers for unstable coarse stationary states (see e.g. \cite{Siet03microscopic}).

For many problems of interest, the selection of the coarse-grained statistics
is often done in an {\em ad hoc} manner, or is based on intuition.
Here we demonstrated the use of simple dynamics arguments and
computations to suggest ``good'' coarse-grained observables,
that are subsequently used in equation-free computation.
For more complex problems, where the appropriate observables for the coarse-grained description
of the system's behavior may be unknown, the coupling of the equation-free approach with
nonlinear data mining approaches such as
Diffusion Maps \cite{Nadl06diffusion,Lafo06diffusion} appears to  be a promising research direction.
This would require the definition of a useful metric quantifying the distance between
nearby graphs (see e.g. \cite{Borg06counting,Vis08graph}).

The computations in this paper were relatively simple, since the coarse description consisted of a single scalar
variable.
For problems where the coarse variables are many (e.g.
for problems where a discretized coarse PDE must be
solved), it is important to note that one does not need
to explicitly estimate each term in large Jacobian matrices.
Matrix-free methods (like Newton-Krylov GMRES) \cite{Saad86gmres:,Kell95iterative}
can be and have been used for the equation-free, time-stepper
based solution of large scale coarse bifurcation problems \cite{Siet03enabling}.
In our discussions of coarse projective integration we demonstrated
a tangible acceleration of the temporal simulation of network evolution.

The computational savings from equation-free methods are obviously very
problem-dependent; for some problems simple direct simulation may be the
easiest way to arrive at, say, a stable stationary state, rather than employing
the equation-free machinery with its associated computational overhead.
There exist, however, tasks such as the location of unstable stationary states, or
the continuation of codimension one and higher bifurcations that would simply be
impossible through direct simulation, yet become accessible through our framework.
As a simple rule of thumb, problems in which there exists a large separation of time
scales between the (slow) evolution of the coarse network evolution
and the (fast) node-level dynamics probably present the greatest potential
for computational savings.
It is, of course, important to also note that, if one is capable of analytically
deriving accurate coarse-grained approximations, the computational savings would be so dramatic
as to obviate the simulations with the detailed model - equation free computations are precisely
intended for situations in which coarse-grained equations are assumed to exist,
but it is not possible to derive them in closed form.

Finally, the crux of the success of the approach lies
almost invariably in the construction of an efficient
{\em lifting algorithm} - a step conceptually
easy to describe, but often problem dependent and
very challenging in itself.
In the case of coarse-graining network dynamics, the problem of constructing networks with prescribed combinations of statistics is a notoriously difficult
one, itself the subject of intense research
\cite{Have55remark,Doro01scaling,Chun02connected,Holm02growing,Klem02growing,Masl02specificity,Volz04random,Serr05tuning,Lear07tuning}.
Given the interdependence of various network statistical properties,
it is also important to be able to efficiently test the graphicality of
a given ``prescription" - is a network with the prescribed set of statistics
even feasible?
We have made this graphicality assumption every time we lifted in our
paper because of the simplicity of our single coarse variable;
yet this is a nontrivial problem when the coarse variables
are more, and interdependent.
In this context, we would also like to mention the
systematic, integer linear programming based approach of
\cite{Goun11generation}.
Clearly, algorithms capable of generating graphs with
prescribed properties can be naturally integrated
in the lifting step of the equation-free framework.

\ack
This work was partially supported by DTRA (HDTRA1-07-1-0005), by the US AFOSR
    and by the US DOE (DE-SC0002097).

\appendix*
\section{A simple singularly-perturbed system of ODEs}

\begin{figure}
\begin{center}
\resizebox{0.63\columnwidth}{!}{%
  \includegraphics{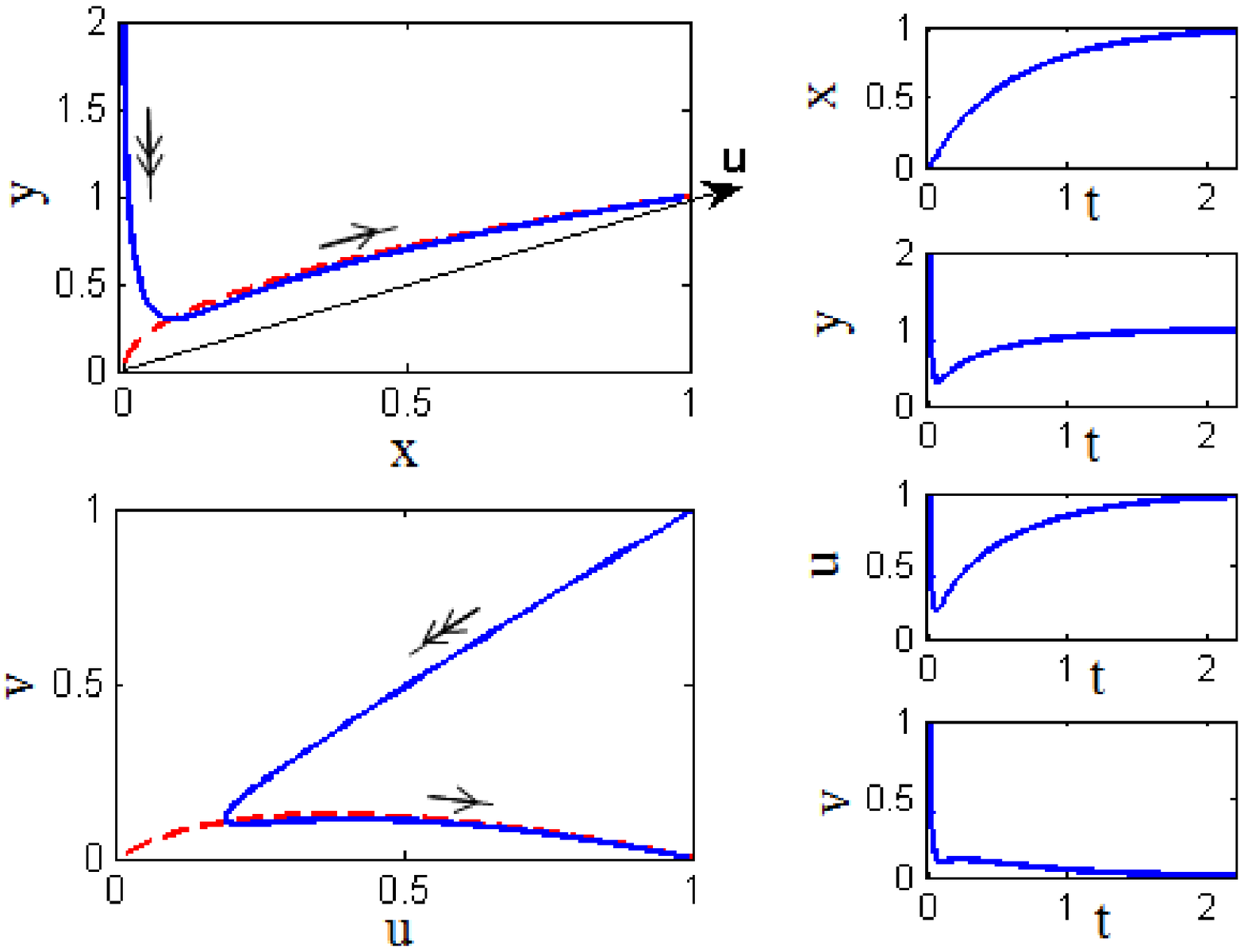}
}
%\vspace{5cm}
\caption{
A simple singularly perturbed dynamical system is used to illustrate the notion of ``pure'' slow variables.
Top-left: Sample trajectory of the dynamical system comprised of Eq.~\ref{eq:SP1} and Eq.~\ref{eq:SP2}
starting from an initial condition $(x,y) = (0,2)$.
Bottom-left: The same trajectory is plotted in terms of coordinates $u = (y+x)/2$ and $v = (y-x)/2$.
In both plots on the left, the slow manifold is shown as a dotted (red) line.
Right: Time series of $x$, $y$, $u$ and $v$ are shown on the right. The label $t$ stands for time.
Notice how in the top case ($x$ is ``pure slow") a short
simulation brings the trajectory to the slow manifold
{\em without practically changing the initial value of $x$}.
In the bottom case ($u$ is not ``pure slow") the short transient again brings the trajectory to the slow manifold, but away from the initial value of $u$ (see also the text).
}
\label{fig:SP}
\end{center}
\end{figure}

Let us consider a simple example of a singularly perturbed system of
ordinary differential equations for illustrative purposes:
\begin{eqnarray}
& \frac{dx}{dt} = 2-x-y,
\label{eq:SP1}
\\
& \frac{dy}{dt} = 50(\sqrt{x}-y).
\label{eq:SP2}
\end{eqnarray}
The top left panel of Fig.~\ref{fig:SP} shows a solution trajectory of this set of equations
starting from the initial condition $(x,y) = (0,2)$ as a solid (blue) line.
The slow manifold of this set of equations, represented by the curve $y = \sqrt{x}$ is plotted
as a dotted (red) line.
The trajectories of $x$ and $y$ plotted on the right show that $x$ initially evolves very
slowly in time (almost not at all), while $y$ has a fast transient approaching the slow manifold, followed by a slow evolution.
It is thus clear from the equations and these figures that $x$ is both a good parameterization
of the slow manifold  (a good coarse variable) and,
at the same time, it is {\em the actual slow variable} of the system: a ``pure" slow variable.

Consider now the same dynamical system, but with a change of coordinates.
Let the new coordinate system be defined by the variables, $u = (y+x)/2$ and $v = (y-x)/2$.
The coordinate $u$ is shown as a slanted arrow in the top left plot of Fig.~\ref{fig:SP}.
It is clear that the variable $u$ is also (similar to $x$) one-to-one with the slow manifold
and hence a suitable candidate for its parameterization,
a ``good" coarse variable.
The solution trajectory computed earlier from the initial condition $(x,y) = (0,2)$ or $(u,v) = (1,1)$
is plotted in terms of coordinates $u$ and $v$ in the lower-left plot of Fig.~\ref{fig:SP}.
The corresponding time series of $u$ and $v$ are plotted to the right.
The system again evolves quickly to the slow manifold,
but $u$ changes quickly in the initial fast transient, away from its initial value.
$u$ has thus both fast and slow components - it is not a
pure slow variable -- even though it can be used to
parameterize the slow manifold.
Additional effort --more than just a few integration steps-- needs to be
invested in finding points that have the initial value
of $u$ {\em and} lie close to/on the slow manifold.

From these figures, it is clear that the direction of fast evolution of the system ($y$ in the original coordinate system,
$(u+v)/2$ in the new coordinate system) has a component along the $u$ coordinate.
Hence, $u$ is not a pure slow variable, but still a good candidate variable
to parameterize the slow manifold: a good coarse variable.

\end{document}